\documentclass{iau}
\usepackage{graphicx}

\title[Optical variability of quasars] 
{Optical variability of quasars: \\ a damped random walk}

\author[\v{Z}eljko Ivezi\'{c} et al.]   
{\v{Z}eljko Ivezi\'{c}$^1$
\and Chelsea MacLeod$^2$
}

\affiliation{
$^1$ Department of Astronomy, University of Washington, \\ Box 351580, Seattle, WA 98195-1580, USA\\ 
                               email: {\tt ivezic@astro.washington.edu} \\[\affilskip]
$^2$  Department of Physics, U. S. Naval Academy, \\ 572c Holloway Rd, Annapolis, MD 21402, USA\\ 
                               email: {\tt macleod@usna.edu} 
}

\pubyear{2014}
\volume{304}  
\pagerange{1--7}
\jname{Multiwavelength AGN Surveys and Studies}
\editors{A. Mickaelian, F. Aharonian \& D. Sanders, eds.}

\begin{document}
\maketitle

\begin{abstract}
A damped random walk is a stochastic process, defined by an exponential covariance matrix
that behaves as a random walk for short time scales and asymptotically achieves a finite variability 
amplitude at long time scales. Over the last few years, it has been demonstrated, mostly but
not exclusively using SDSS data, that a damped random walk model provides a satisfactory statistical 
description of observed quasar variability in the optical wavelength range, for rest-frame timescales from 
5 days to 2000  days. The best-fit characteristic timescale and asymptotic variability amplitude 
scale with the luminosity, black hole mass, and rest wavelength, and appear independent of redshift.
In addition to providing insights into the physics of quasar variability, the best-fit model 
parameters can be used to efficiently separate quasars from stars in imaging surveys with 
adequate long-term multi-epoch data, such as expected from LSST. 
\end{abstract}

\firstsection 

\section{Introduction}

Quasars are variable sources with optical amplitudes of several tenths of a magnitude for time scales 
longer than a few months (e.g., Hawkins \& Veron 1995; Trevese et al. 2001;
Ivezi\'{c} et al. 2004; Vanden Berk et al. 2004). Sesar et al. (2007) and Butler \& Bloom (2011) showed using 
SDSS Stripe 82 data (a $\sim$300 deg$^2$ equatorial region imaged about 60 times) that practically all 
quasars spectroscopically confirmed by SDSS are photometrically variable. 

Quantitative statistical description of quasar variability is important both for understanding the physics of 
the driving mechanism(s), and for selecting quasars in imaging surveys. Here we describe recent progress in the
analysis of quasar variability which demonstrated that a stochastic process called damped random 
walk (DRW) provides a satisfactory statistical description of quasar variability in the optical wavelength range.

\section{Quantitative analysis of quasar variability}

Two main methods have been utilized over the last few decades to quantitatively describe stochastic 
quasar variability: a variability structure function analysis and direct modeling of light curves. 

\subsection{Structure function approach} 

The structure function as a function of time lag $\Delta t$, SF($\Delta t$), is equal to the standard deviation 
of the distribution of the magnitude difference $m(t_2)-m(t_1)$ evaluated at many different times $t_1$ and 
$t_2$, such that time lag $\Delta t = t_2-t_1$ (and divided by $\sqrt{2}$ because of differencing; we warn the 
reader that a number of slightly varying definitions have been used in recent quasar studies). This operational 
definition has been applied to both light curves of individual objects and as an ensemble analysis tool that is 
applicable even when only two photometric measurements per object are available (in this case it is typically assumed 
that quasars selected from narrow luminosity and redshift bins have statistically the same variability behavior). 

The time dependence of the structure function was found to be consistent with the prediction based on a damped 
random walk model (MacLeod et al. 2012): 
\begin{equation}
        {\rm SF}(\Delta t) = {\rm SF}_\infty \, \left[1 - \exp(- \Delta t / \tau) \right]^{1/2}
\end{equation} 
(for illustration see Figure 2 in the contribution by Ivezi\'{c} et al. in these Proceedings). 
At small time lags, ${\rm SF}(\Delta t) \propto \Delta t ^ {1/2}$, and thus a DRW is equivalent to 
an ordinary random walk for $\Delta t \ll \tau$ (the ``damped'' aspect manifests itself as 
SF($\Delta t) \rightarrow$ SF$_\infty$ for $\Delta t \gg \tau$). 

\begin{figure}[t]
\vskip -1.7in
\begin{center}
 \includegraphics[width=0.99\textwidth]{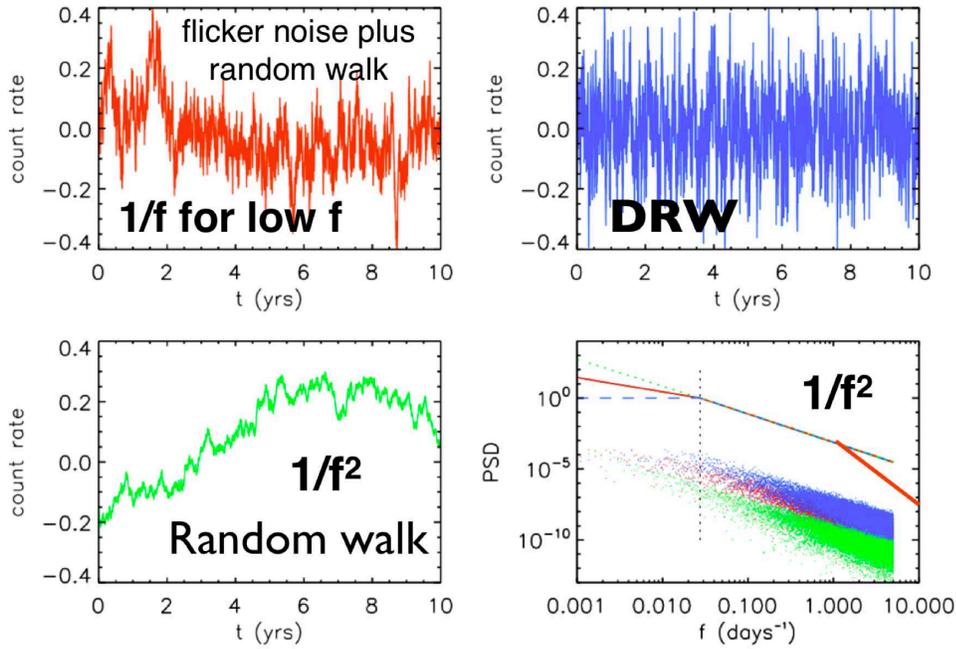} 
\vskip -1.7in
 \caption{A comparison of simulated light curves generated using three different power spectral density functions (PSD),
which are illustrated in the bottom right panel by lines (solid line: top left panel; dashed line: top right panel;
dotted line: bottom left panel). In all three cases, the PSD at short time scales (large frequency $f$) is 
proportional to $f^{-2}$ (the PSD for a random walk has the same index at {\it all} frequencies). The transition time scale is 
given by $\tau=5.8$ days, corresponding to transition frequency $f_t = (2\pi\tau)^{-1}$ (shown by the vertical dotted line
in the bottom right panel). The PSD at long time scales ($f < f_t$) follows $f^\alpha$, with $\alpha=-1$ (top 
left panel, a mixture of a flicker noise and a random walk), $\alpha=-1.9$ (bottom left, almost identical to a random walk) and 
$\alpha=0$ (top right, similar to a DRW). The y axis in the bottom right panel is in arbitrary units. Observed 
light curves of quasars are consistent with $-1 < \alpha < 0$, while $\alpha < -1$ is ruled out by SDSS Stripe 82 data.
The solid line at $f>1$ in the bottom right panel illustrates departures from the $f^{-2}$ PSD at the shortest timescales 
found using Kepler data (Mushotzky et al. 2011). Adapted from MacLeod et al. (2010).}
\label{fig:PSDs}
\end{center}
\end{figure}

The structure function is related to the autocorrelation function, which makes a Fourier pair with 
the power spectral density function (PSD). The PSD for a DRW is given by 
\begin{equation}
                 {\rm PSD}(f) = { \tau^2 \, {\rm SF}^2_\infty \over  1 + (2\pi f \tau)^2 }.
\end{equation}
Therefore, a DRW is a $1/f^2$ process at high frequencies, just as an ordinary random walk
(when ${\rm SF} \propto (\Delta t)^\gamma$, then ${\rm PSD} \propto 1/f^{(1+2\gamma)}$). 
The ``damped'' nature is seen as a flat PSD at low frequencies ($f\ll 2\pi/\tau$).
A comparison of light curves drawn from a DRW and two other
stochastic processes with similar PSDs is shown in Figure~\ref{fig:PSDs}.

\subsection{Direct modeling of light curves as a damped random walk} 

Observed light curves can be used to directly constrain the DRW model parameters, $\tau$ and
SF$_\infty$ (Kelly, Bechtold \& Siemiginowska 2009, hereafter KBS09; Koz{\l}owski et al. 2010, MacLeod 
et al. 2010, 2011, 2012; Zu et al. 2012). Before summarizing the main results, we briefly review the
statistical properties of a DRW. 

The CAR(1) process, as it is called in statistics literature, for a time series $m(t)$ is described by a 
stochastic differential equation which
includes a damping term that pushes $m(t)$ back to its mean (see KBS09). Hence, it is also known 
as a DRW in astronomical literature (another often-used name is the Ornstein--Uhlenbeck 
process, especially in the context of Brownian motion). In analogy with calling a random walk 
a ``drunkard's walk,''  a DRW could be justifiably called a ``married drunkard's walk'' 
-- who always comes home to his or her spouse instead of drifting away.

Stochastic light curves can be modeled using the covariance matrix. For a DRW, the covariance
matrix is 
\begin{equation}
\label{eq:DRWSij}
              S_{ij}(\Delta t_{ij}) = \sigma^2 \exp(-|\Delta t_{ij}| / \tau),
\end{equation}
where $\Delta t_{ij} = t_i - t_j$, and $\sigma$ and $\tau$ are model parameters; $\sigma^2$ controls 
the short timescale covariance ($\Delta t_{ij} \ll \tau$), which decays exponentially on a timescale
given by $\tau$. The corresponding autocorrelation function is ${\rm ACF}(t) = \exp(-t / \tau)$. The
asymptotic value of the structure function, SF$_\infty$, is equal to $2\sigma$. 
A number of other convenient models and parametrizations for the covariance matrix are discussed 
in Zu et al. (2012).

\subsection{Tests of a damped random walk model} 

Both a structure function analysis and the direct modeling of light curves demonstrate that a DRW 
provides a good description of the optical continuum variability of quasars. For example, 
the time span of SDSS data from Stripe 82 is sufficiently long to constrain $\tau$ for the majority of 
the $\sim$10,000 quasars with light curves (MacLeod et al. 2010, 2011). The best-fit values of 
$\tau$ and SF$_\infty$ are correlated with physical parameters, as discussed in the next section. 

MacLeod et al. (2010) have concluded that the observed light curves of quasars are consistent 
with PSD$\propto f^\alpha$ at long timescales, with $-1 < \alpha < 0$, while $\alpha < -1$ is ruled 
out. Furthermore, Zu et al. (2012) have analyzed
OGLE light curves using a number of stochastic processes with covariance matrices similar to that
for a DRW. They concluded that the DRW model is consistent with data on the probed time scales
(from a month to a few years). 
Some deviations from the DRW model were detected at short timescales (a month or less) by 
Mushotzky et al. (2011) using high-precision Kepler data. They found that the measured PSD 
at high frequencies (from $10^{-6}$ Hz up to $10^{-5}$ Hz) is steeper than the expected $f^{-2}$ 
behavior. 

The distribution of magnitude differences drawn from a DRW light curve should be Gaussian. 
The number of points per observed light curve is typically too small to test this expectation 
using individual objects. When using an ensemble analysis, the observed distribution is puzzlingly 
closer to an exponential (Laplace) distribution than to a Gaussian distribution (Ivezi\'{c} et al. 2004;
MacLeod et al. 2008). Nevertheless, MacLeod et al. (2012) showed that the exponential distributions 
seen in the statistics of ensembles of quasars naturally result from averaging over quasars that are 
individually well described by a Gaussian DRW process.

\vskip -0.3in \phantom{x}
\section{Insights into the physics of quasar variability}

The best-fit values of $\tau$ and SF$_\infty$, determined using a DRW model and SDSS Stripe 82 
light curves, are correlated with physical parameters, such as the luminosity, black hole mass, and 
rest-frame wavelength (MacLeod et al. 2010, 2012). Their analysis shows SF$_\infty$ to increase with 
decreasing luminosity and rest-frame wavelength (as was observed previously), and without a 
correlation with redshift. They found a correlation between SF$_\infty$ and black hole mass with a 
power-law index of $0.18\pm0.03$, independent of the anti-correlation with luminosity. They
also found that $\tau$ increases with increasing wavelength with a power-law index of 0.17, remains 
nearly constant with redshift and luminosity, and increases with increasing black hole mass with a 
power-law index of $0.21\pm0.07$. 

The amplitude of variability is anti-correlated with the Eddington ratio, which suggests a scenario 
where the optical fluctuations are tied to variations in the accretion rate, possibly in an inhomogeneous
accretion disk (Dexter \& Agol 2011). However, an additional dependence on luminosity and/or black 
hole mass was found that cannot be explained by the trend with Eddington ratio.  Recent studies 
show evidence for enhanced color variability compared to what is expected if the mean 
accretion rate is solely driving the variability (Schmidt et al. 2012), which is consistent with a scenario 
involving hot spots in the disk. 

\vskip -0.3in \phantom{x}
\section{Conclusions} 

The last decade has seen enormous progress in both data availability and the modeling of stochastic
quasar variability. The damped random walk model provides a satisfactory statistical description for 
practically all the data available at this time. This progress is likely to continue thanks to new post-SDSS
massive sky surveys. For example, the Large Synoptic Survey Telescope (LSST; for a brief overview 
see Ivezi\'{c} et al. 2008) will extend the light curve baseline for $\sim$10,000 quasars from SDSS Stripe 82 
to over 30 years, and will obtain an additional $\sim$800 high-precision ($\sim$0.01 mag) photometric 
measurements. In particular, these data will enable a definitive robust measurement of the low-frequency 
behavior of the PSD for quasar variability.

\end{document}